\documentclass[a4paper]{spie}  %>>> use this instead for A4 paper
\usepackage{amsmath,amsfonts,amssymb}
\usepackage{graphicx}
\usepackage[colorlinks=true, allcolors=blue]{hyperref}

\title{VLTI images of circumbinary disks around evolved stars}

\author[a]{Jacques Kluska}
\author[a]{Rik Claes}
\author[a]{Akke Corporaal}
\author[a]{Hans Van Winckel}
\author[b]{Javier Alcolea}
\author[c]{Narsireddy Anugu}
\author[d]{Jean-Philippe Berger}
\author[a]{Dylan Bollen}
\author[b]{Valentin Bujarrabal}
\author[e]{Robert Izzard}
\author[f]{Devika Kamath}
\author[g]{Stefan Kraus}
\author[d]{Jean-Baptiste Le Bouquin}
\author[h]{Michiel Min}
\author[i]{John D. Monnier}
\author[j]{Hans Olofsson}

\affil[a]{KULeuven, Celestijnenlaan 200D, 3001 Leuven (Belgium)}
\affil[b]{Observatorio Astronómico Nacional (Spain)}
\affil[c]{University of Arizone (United States)}
\affil[d]{Institut de Planétologie et d’Astrophysique de Grenoble (France)}
\affil[e]{University of Surrey (UK)}
\affil[f]{Macquarie University (Australia)}
\affil[g]{University of Exeter (UK)}
\affil[h]{SRON Netherlands Institute for Space Research (Netherlands)}
\affil[i]{Univ. of Michigan (United States)}
\affil[j]{Chalmers University of Technology (Sweden)}

\authorinfo{Further author information: (Send correspondence to Jacques Kluska)\\Jacques Kluska: E-mail: jacques.kluska@kuleuven.be, Telephone: (+32) 16 32 84 51\\}

\begin{document} 
\maketitle

\begin{abstract}
The new generation of VLTI instruments (GRAVITY, MATISSE) aims to produce routinely interferometric images to uncover the morphological complexity of different objects at high angular resolution. Image reconstruction is, however, not a fully automated process. 
Here we focus on a specific science case, namely the complex circumbinary environments of a subset of evolved binaries, for which interferometric imaging provides the spatial resolution required to resolve the immediate circumbinary environment.
 Indeed, many binaries where the main star is in the post-asymptotic giant branch (post-AGB) phase are surrounded by circumbinary disks. Those disks were first inferred from the infrared excess produced by dust. Snapshot interferometric observations in the infrared confirmed disk-like morphology and revealed high spatial complexity of the emission that the use of geometrical models could not recover without being strongly biased. Arguably, the most convincing proof of the disk-like shape of the circumbinary environment came from the first interferometric image of such a system (IRAS08544-4431) using the PIONIER instrument at the VLTI. This image was obtained using the SPARCO image reconstruction approach that enables to subtract a model of a component of the image and reconstruct an image of its environment only. In the case of IRAS08544-4431, the model involved a binary and the image of the remaining signal revealed several unexpected features. Then, a second image revealed a different but also complex circumstellar morphology around HD101584 that was well studied by ALMA. To exploit the VLTI imaging capability to understand these targets, we started a large programme at the VLTI to image post-AGB binary systems using both PIONIER and GRAVITY instruments.
\end{abstract}


\section{INTRODUCTION}
\label{sec:intro}  % \label{} allows reference to this section
A significant fraction of stars are binaries and binary research constitutes a main domain of stellar astrophysics. 
Binary evolution gives birth to diverse phenomena such as thermonuclear novae, supernovae type Ia, sub-luminous supernovae, gravitational waves and objects such as subdwarf B-stars, barium stars, cataclysmic variables, and asymmetric planetary nebulae.
However, binary evolution is not yet well understood.
Here, we focus on a peculiar stage of binary evolution that are the post-asymptotic giant branch (post-AGB) binaries.
During the preceding phase, the AGB phase, the star has strong winds that expel its atmosphere.
In the case of a close binary (with a separation less than few astronomical units) this interaction produces a circumbinary disk that is observable in the post-AGB phase through the infrared excess in the spectral energy distribution\cite{VanWinckel2003,Gezer2015} reminiscent of young stellar objects.
As the binary nature is difficult to reveal during the AGB phase, because of stellar pulsations and atmosphere motions that blur the radial velocity signal but also because of the dense dusty surroundings of the star, the post-AGB phase is ideal to probe binarity and the system surroundings as the atmosphere of the evolved star is almost all expelled and the direct environment is not obscuring the stars anymore.
We can, therefore, detect and characterise the orbits of post-AGB binary systems\cite{Oomen2018}.
However, their periods and eccentricities do not correspond to the expected ones from binary interaction models\cite{Nie2012}.

It is now well known that binarity and the presence of a circumbinary disk are tightly linked\cite{VanWinckel2018}.
Disk-binary interaction can therefore be the key to solve the discrepancy between theory and observations.
Those disks were further studied by millimetre interferometry showing their stability as they are in Keplerian rotation\cite{Bujarrabal2013,Bujarrabal2015,Bujarrabal2016,Bujarrabal2017,Bujarrabal2018}.
However, the angular scale achieved with those observations do not allow to probe the inner disk regions where the disk-binary interaction processes happen.

To explore this, one needs to study the inner disk structure.
As the distances to those objects are relatively large ($\sim$1\,kpc), those disks appear compact on the sky and the use infrared interferometry is necessary to achieve the needed angular resolution power to spatially resolve them.
In this paper we show how infrared interferometry offers unique capabilities to constrain the properties of the systems and how the interferometric imaging technique contributed to advance our knowledge on these peculiar objects.

In this paper, we will describe the first snapshot-like studies and surveys using infrared interferometers with sparse (u, v)-coverage (Sect.\,\ref{sec:snapshot}). Then we will show what interferometric imaging brings to the understanding of these systems following two examples: IRAS08544-4431 (Sect.\,\ref{sec:IRAS08}) and HD101584 (Sect.\,\ref{sec:HD101}). We will finish by presenting INSPIRING: an interferometric imaging large programme at the VLTI on post-AGB binaries (Sect.\,\ref{sec:INSPIRING}).

\section{Interferometric snapshots}
\label{sec:snapshot}

First infrared interferometric snapshot observations were performed using the MIDI and AMBER instruments on the VLTI strengthening the idea that the circumbinary environment has a compact disk-like structure\cite{Deroo2006,Deroo2007}.
Further observations using VLTI and CHARA interferometric instruments confirmed it as well and allowed a successful in-depth modelling of both the SED and the interferometric data using radiative transfer models of protoplanetary disks\cite{Hillen2013,Hillen2014,Hillen2015}.
One of those models based on AMBER and MIDI data allowed to localise the PAH emission observed in a unique target, HR4049, to be in the outflow\cite{Acke2013}.

In order to confirm the first conclusions from observations of individual targets, interferometric surveys were carried out subsequently. 
First, in the mid-infrared\cite{Hillen2017} using MIDI, a survey of 19 objects have showed a strong similarity between the observations and a synthetic population of models of protoplanetary disks, demonstrating that the disk morphology is best suited to reproduce both the interferometric observations and the SED for a larger sample of targets.
Moreover, a similar mid-infrared size-color relation was found for those disks as for disks around young stellar objects\cite{Menu2015}.

A similar snapshot survey was obtained in the near-infrared on a sample of 23 sources\cite{Kluska2019} using the PIONIER\cite{LeBouquin2011} instrument on the VLTI.
Using a set of geometrical models with increasing complexity it was possible to show that the near-infrared size/stellar luminosity relation found for young stellar objects\cite{Monnier_2005,Lazareff2017}, can be extrapolated to the high-luminosities that are typical for post-AGB stars ($10^3-10^4$\,L$_\odot$).
This demonstrates that the inner dusty rims of disks around post-AGB binaries are not ruled by dynamical truncation due to the inner binary but by dust sublimation.
Those near-infrared sizes of the environment are, however, slightly larger than in protoplanetary disks likely because of different dust chemistry and/or gas density in the dust sublimation region.
This is corroborated by the slightly lower environmental temperatures measured using the spectral channels of the PIONIER instrument.
The other main conclusion of this survey was that complex geometric models (more than 13 parameters) are favoured for most ($>$55\%) targets.
This suggests a high intrinsic complexity of the target morphologies in the near-infrared.
There is, therefore, a risk of being model biased when trying to reproduce the complex interferometric signal provided by 4-telescope infrared interferometric instruments.
With the advent of instruments being able to combine more and more telescopes in the infrared, interferometric image reconstruction techniques are perfect to solve this problem and to reveal the origin of the complexity of these targets.

\section{First image: IRAS08544-4431}
\label{sec:IRAS08}

\begin{figure*}[t]
    \includegraphics[width=5.7cm]{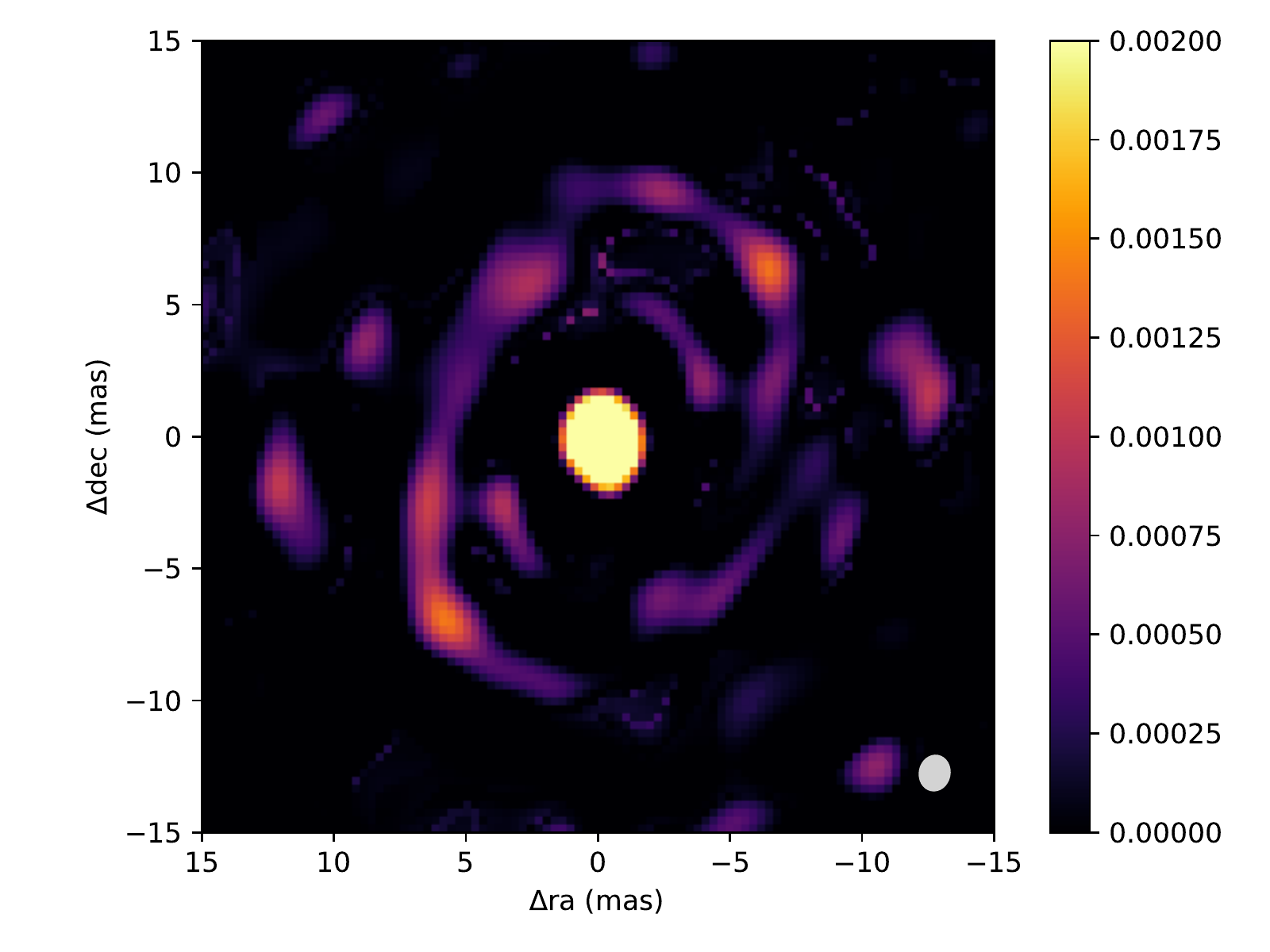}
    \includegraphics[width=5.7cm]{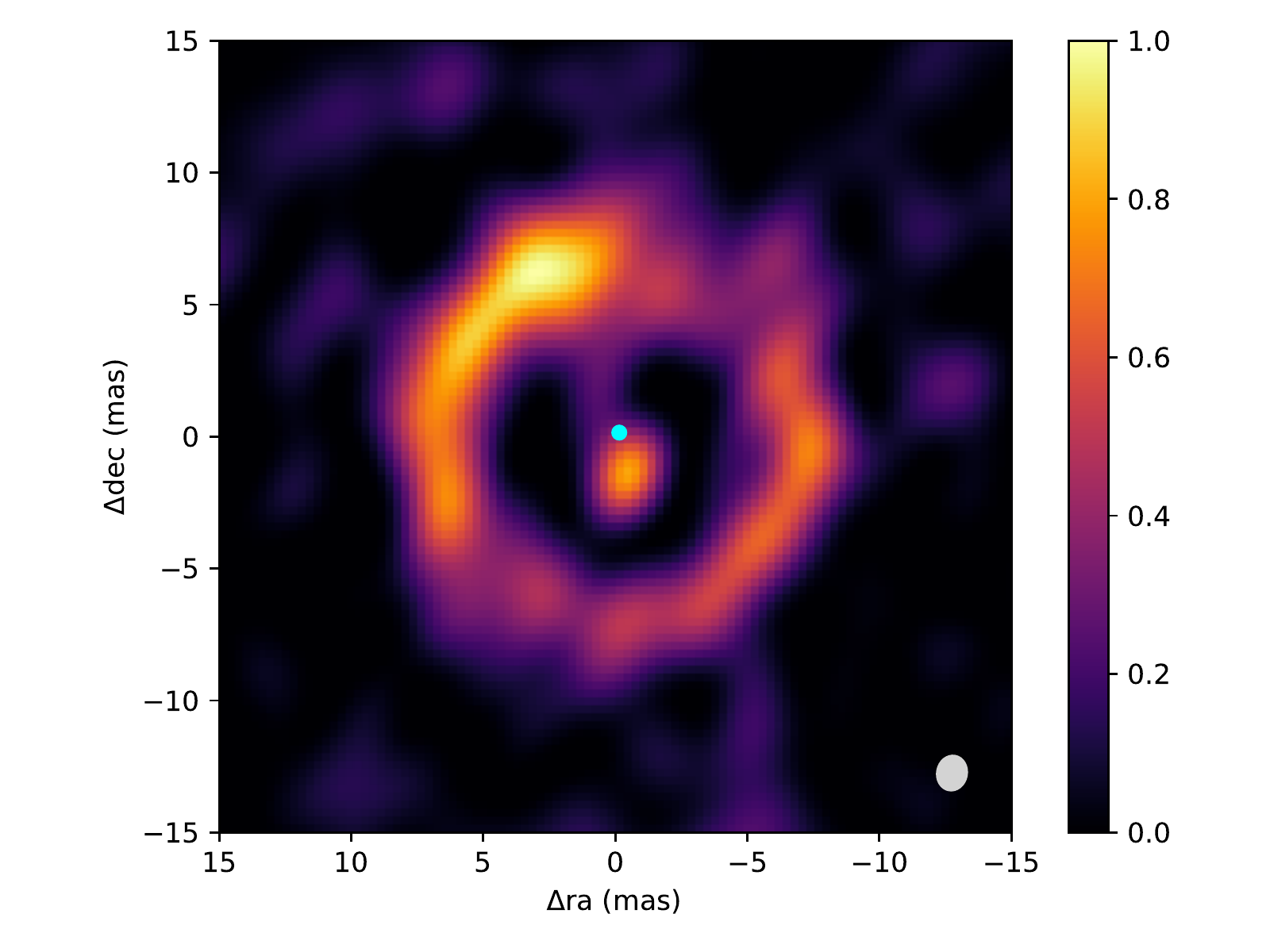}
    \includegraphics[width=5.7cm]{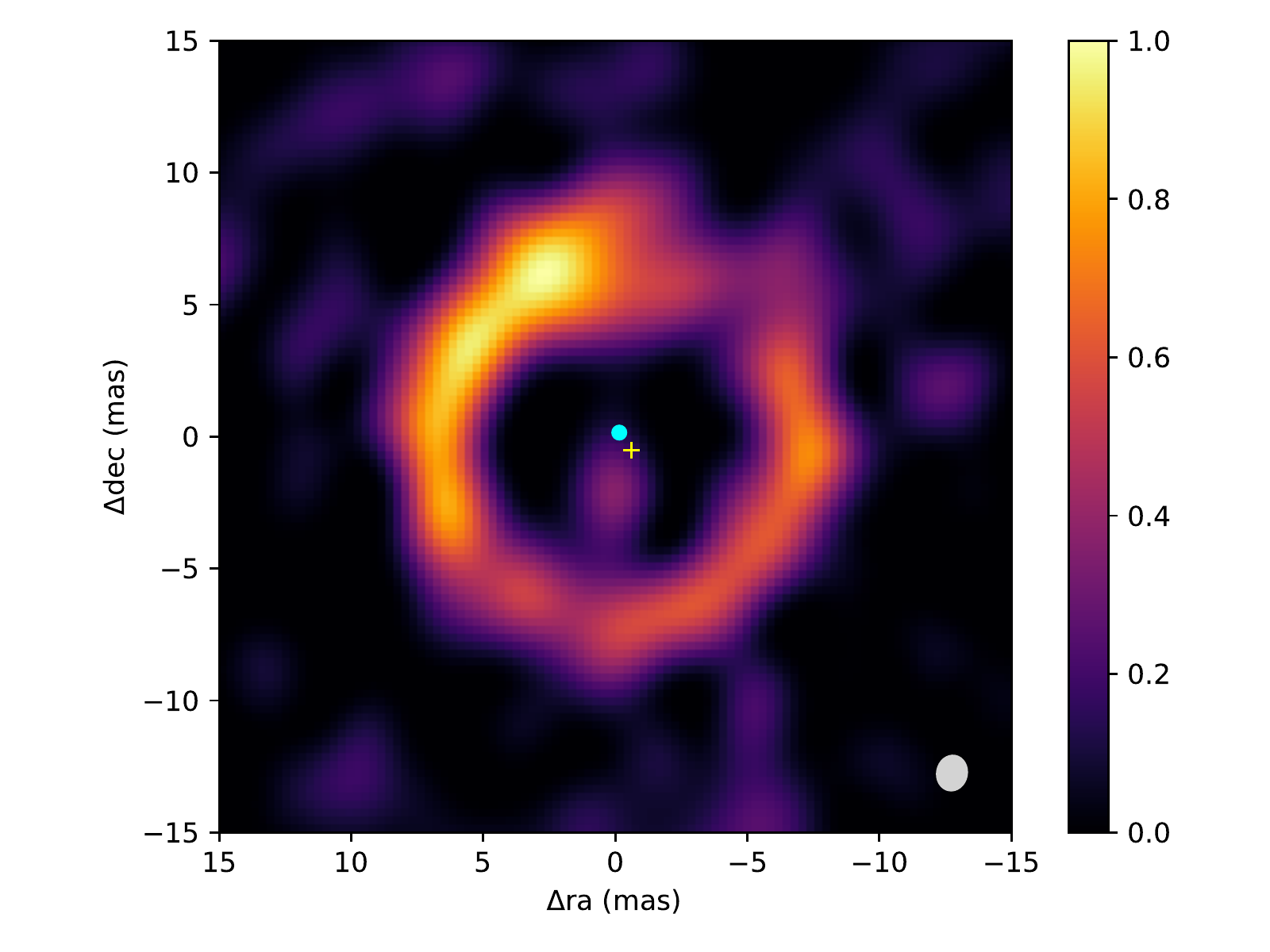}
    \caption{Image reconstructions of IRAS08544-4431 from VLTI/PIONIER data\cite{Hillen2016,Kluska2018}. Left: monochromatic image reconstruction. The flux level was cut to better see the environment of the star. Center: SPARCO/MiRA image reconstruction subtracting a single star with diameter 0.5\,mas, $f^*_0$=61.0\% and $d_\mathrm{env}$=0.182. This image has revealed emission coming from the secondary. Right: SPARCO/MiRA image reconstruction subtracting the primary star and contribution from the secondary, revealing the circumbinary disk structure. The cyan disk indicates the position and size of the primary star. The yellow cross indicates the location of the emission from the secondary star. The beam size is indicated in the bottom-right corner.}
    \label{fig:IRAS08}
\end{figure*}

The first interferometric image in the infrared was performed using the VLTI/PIONIER instrument on IRAS08544-4431 and revealed a very complex structure\cite{Hillen2016} as can be seen in Fig.\,\ref{fig:IRAS08}.
To be able to retrieve the image of the circumstellar environment, the Semi-Parametric Approach for Reconstruction of Chromatic Objects (SPARCO) proved to be crucial, as it assigns a geometrical model and a spectrum to the central star and removes it from the image, allowing a dramatic increase of the quality of the reconstructed image of the environment\cite{Kluska2014}.
This technique was successfully applied to image the environments of young stellar objects\cite{Kluska2016,Kluska2020a} and is particularly adapted to circumstellar environments.

As a first step we have modelled the star as uniform disk with a diameter of 0.5\,mas as predicted from the SED modeling.
We also fitted the stellar-to-total flux ratio ($f^*_0$) to be $\sim$61\% of the total flux at 1.65$\mu$m and the spectral index ($d_\mathrm{env}$) to be $\sim$0.182 (defined as $d_\mathrm{env}=\frac{\mathrm{d}\log F_\lambda}{\mathrm{d}\log\lambda}$, and that corresponds in that case to a temperature of $\sim$1700\,K).
In the resulting image we can see a clear ring but also a point source emission close to the primary.
This emission is likely associated with the secondary star.
It was unexpected as the secondary is star is likely a main-sequence star and, hence, the contrast should be such that it is not detectable in our observations. 
As we explain it later on, we interpreted this emission as likely coming from a circum-secondary accretion disk.

We therefore used SPARCO to subtract the emission from the position of the secondary as well.
We used a geometrical model to find the best fitting positions and flux contribution of the secondary and use them in the SPARCO image reconstruction.
The resulting image provides a better quality reconstruction of the inner disk rim.
The disk is seen almost pole-on and has complex azimuthal structures showing two maxima and two minima.

We then modelled the data using the radiative transfer code MCMax that was developped for modeling protoplanetary disks\cite{Kluska2018}.
The model is able to reproduce both the SED and the squared visibilities from PIONIER.
However, given that the model is axi-symmetric, it was not possible to reproduce the closure phase signal.
This confirms that the structure we see in the image reconstruction has an astrophysical origin.
Moreover, there is a large amount ($\sim$15\%) of over-resolved flux (the visibility level at the shortest baseline is below one) that could not be reproduced by the radiative transfer model alone.

Finally, this imaging experience allowed us to build an archetype of the inner regions of these post-AGB binaries with the following building blocks:
\begin{itemize}
    \item the post-AGB star that emits $\sim$60\% of the total flux in $H$-band.
    \item the inner rim of the circumbinary disk is at the dust sublimation radius and is azimuthally perturbed
    \item an unexpected emission from the secondary main-sequence star. This secondary star should not be detectable given the expected contrast with the post-AGB star. We hypothesised that the emission comes from the circum-secondary accretion disk. This is supported by the systematic detection of jets originating from the secondary star in H$_\alpha$ time-series spectra\cite{Gorlova2015,Bollen2017,Bollen2019,Bollen2020}.
    \item A strong contribution from an over-resolved emission is detected. The origin of this emission is unknown yet and cannot be reproduced by radiative transfer model including a disk alone.
\end{itemize}
Interferometric imaging was crucial to reveal such a detailed picture of this complex system which is an archetype for a large number of evolved binaries. 
However, it is yet unclear if IRAS08544-4431 is a typical example of post-AGB binary systems.
One needs to study other targets of this class.

\section{Revealing a complex environment: imaging HD101584}
\label{sec:HD101}

\begin{figure*}[t]
\centering
    \includegraphics[width = 12cm]{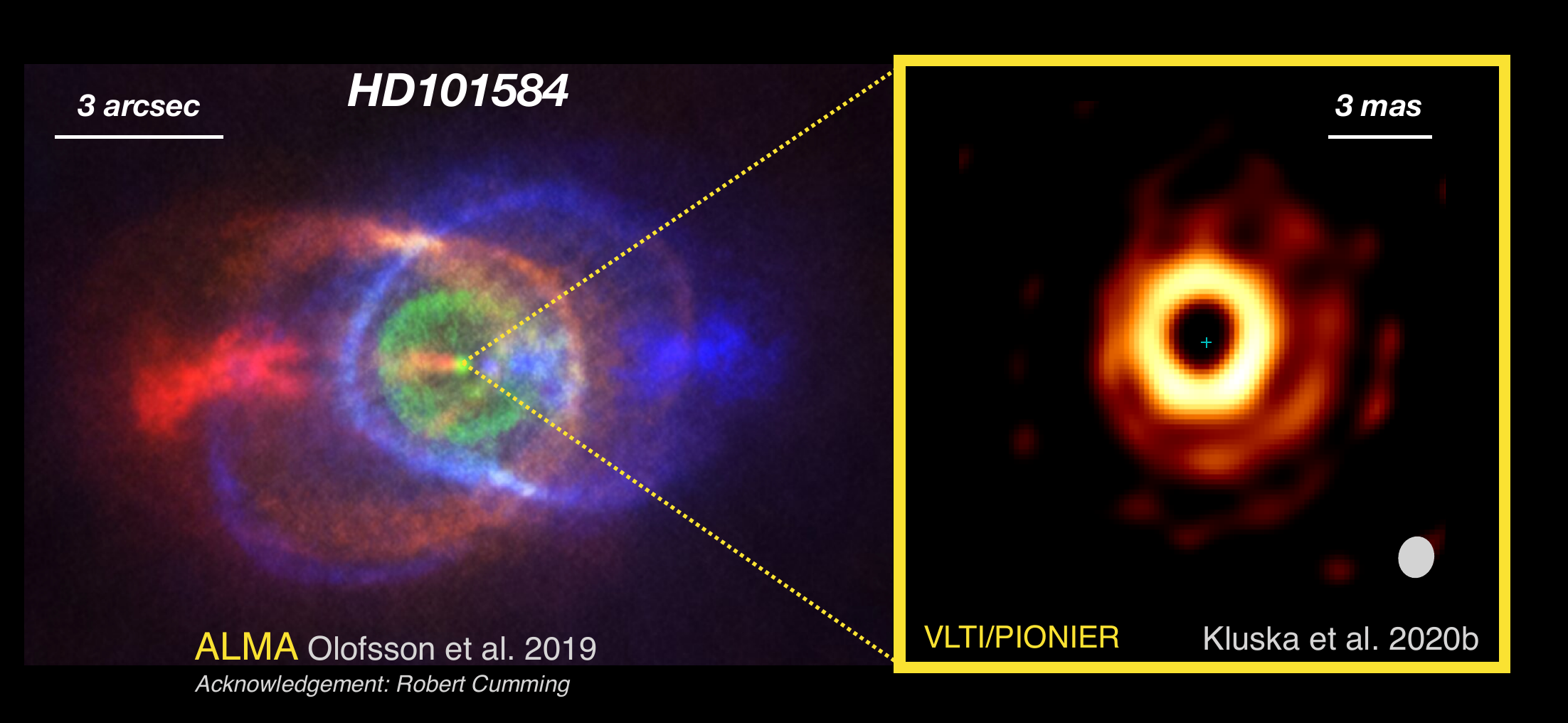}
    \caption{Two physical scales probed by ALMA and PIONIER. Left: ALMA image in CO(2–1) with a resolution of 85\,mas. Right: VLTI/PIONIER image with a 2\,mas resolution. The star, that is subtracted from the image using SPARCO, is indicated by the green cross. The beam of the VLTI observations is in the bottom-right corner. }
    \label{fig:HD101584ALMA}
\end{figure*}

\begin{figure*}[t]
    \includegraphics[width=5.7cm]{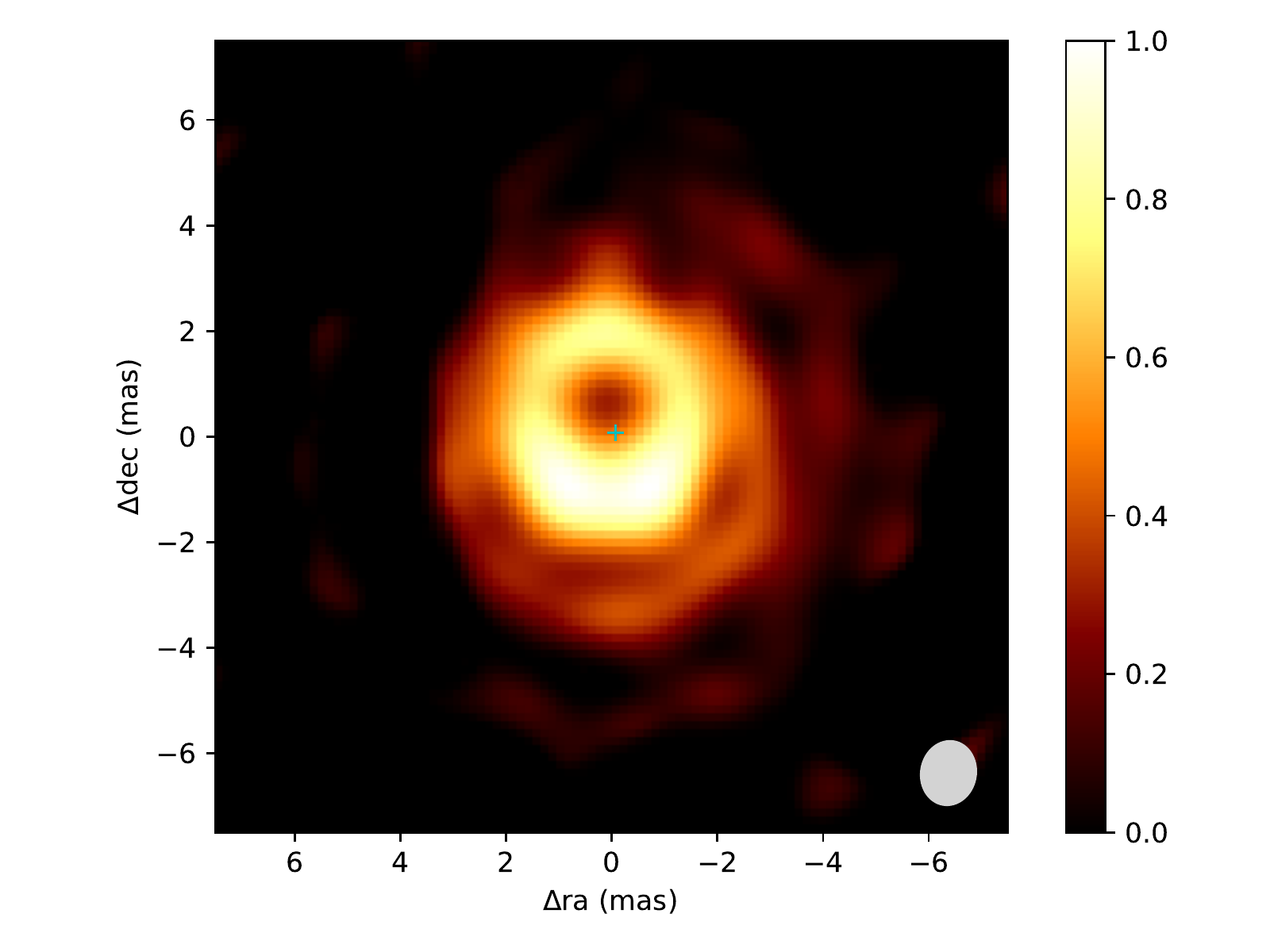}
    \includegraphics[width=5.7cm]{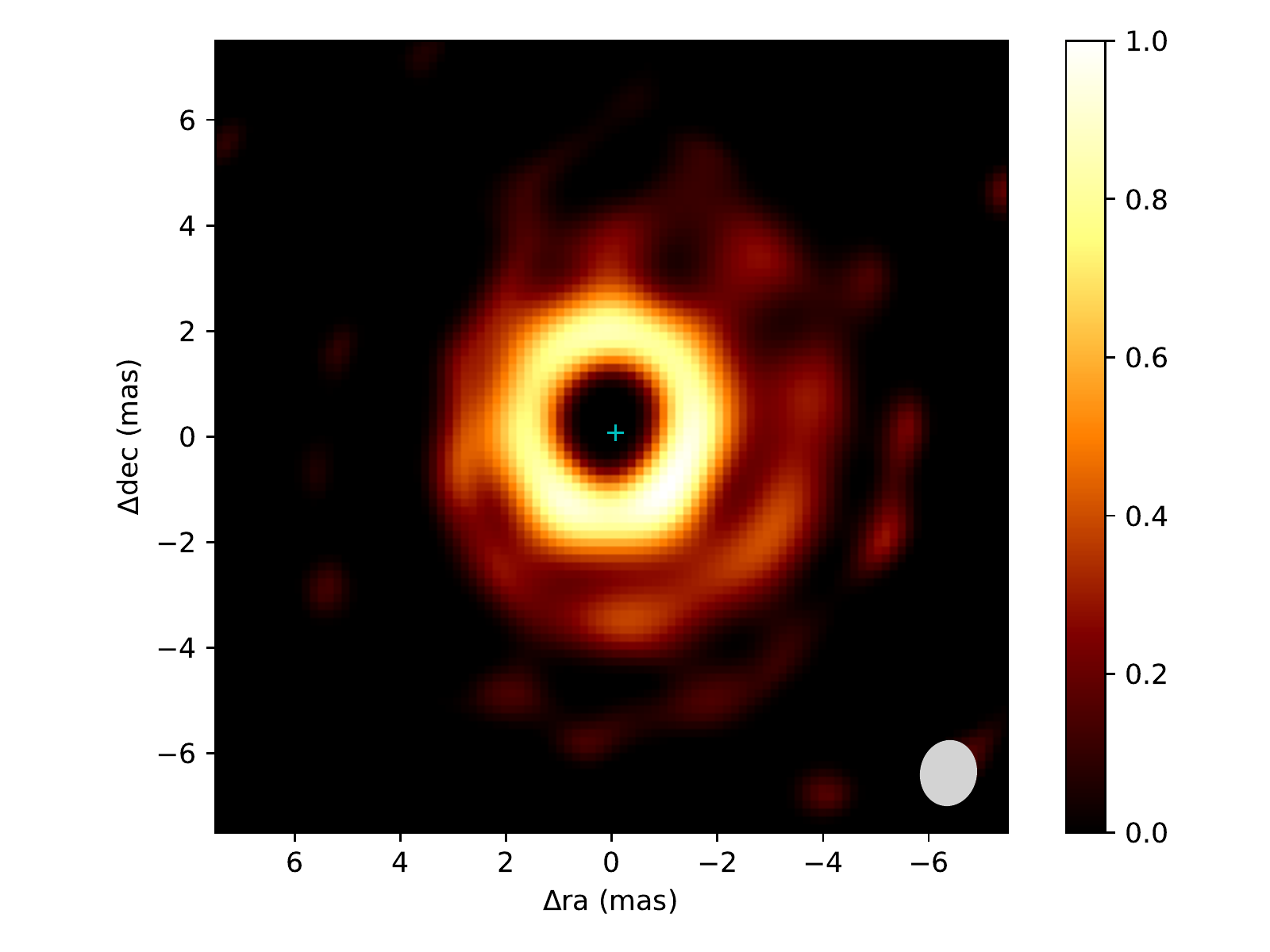}
    \includegraphics[width=5.7cm]{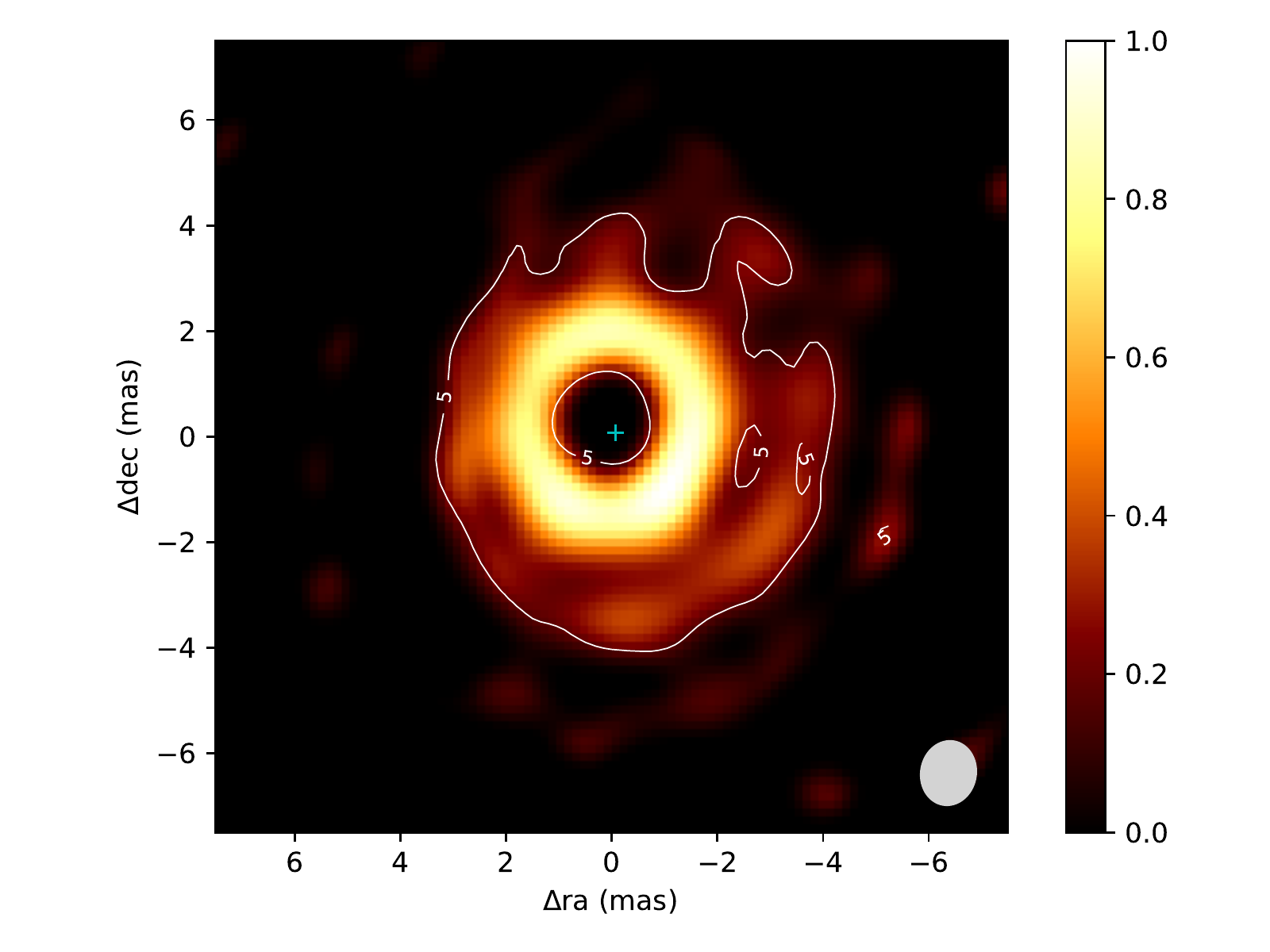}
    \caption{VLTI/PIONIER image reconstructions of HD101584 using MiRA/SPARCO\cite{Kluska2020b}. Left: image with $f^*_0$=24.4\% and $d_\mathrm{env}$ = 2.45. Center and right: image with $f^*_0$=28.0\% and $d_\mathrm{env}$ = -1.75. On the right image the contours indicate the 5-$\sigma$ significance of the emission as estimated by the bootstrapping.}
    \label{fig:HD101584}
\end{figure*}

Another infrared interferometric imaging campaign was performed on HD101584 that has been extensively studied with the Atacama Large Millimeter/submillimeter Array (ALMA)\cite{Olofsson2015,Olofsson2017,Olofsson2019}.
These observations revealed a bipolar environment as traced by the CO gas (see Fig.\,\ref{fig:HD101584ALMA}).
An hourglass structure is dominating the large scale morphology with high-velocity outflows.
Close to the star, there is a disk-like structure seen both in CO and in the continuum.
An environment with such a symmetry axis is likely to be the product of a strong binary interaction.
However, photometric\cite{Bakker1996} and spectroscopic\cite{Diaz2007} time-series do find a periodic signal but with different and contradicting periods (218\,days and 144\,days resp.) and the detections do not seem to be significant enough.
Today's binary status of the target is therefore unclear, but it is likely that, if not binary today, the object has been a binary in the past and merged.

The observations of the inner regions of this target were made with VLTI/PIONIER.
We used the SPARCO technique again to recover the image from the interferometric dataset.
Here, we subtracted a single point-source and retrieved a stellar-to-total flux ratio of $f^*_0$ of 24.4\% and $d_\mathrm{env}$ of 2.45.
The image shows a complex morphology that can be divided into an inner and an outer morphology (see Fig.\,\ref{fig:HD101584}).
The inner morphology appears as a ring.
The outer morphology seems also to form a ring-like feature even though the emission is not azimuthally continuous.
This second ring seems to be shifted compared to the first ring.
No secondary is detected. It could be due to the lack of angular resolution or because the target is a merger \cite{Kluska2020b}.
The rest of the emission is not significant as estimated from bootstrapping.

This first image inspired the geometrical models that include two rings, a central star and a background emission.
The geometrical models have shown that the inner ring has a temperature of 1540$\pm$10\,K while the outer one is colder with a temperature of 1014$\pm$10\,K.
This geometrical model confirmed the shift between the two rings with a fitted value of 0.57$\pm$0.01\,mas.
It is interesting to note that the shift is in the direction of the minor axis as it would be the case for an emission above the disk mid-plane.
The stellar-to-total flux ratio obtained from the geometrical model is 28\%.
As a cross-check, we have used this stellar-to-total flux ratio to produce a new image reconstruction.
We can see that the regions around the star are depleted from emission as expected with a higher $f^*_0$ (see Fig.\,\ref{fig:HD101584}).

Several scenarios could explain the results from interferometric imaging and geometrical model fitting (see the discussion in \cite{Kluska2020b}) but here is what we consider the most likely scenario.
The inner ring size (4.1\,au in radius) is compatible with the dust sublimation rim of the circumstellar disk.
The outer ring does not lie in the plane of the disk and is above at a height of 1.6\,au.
We speculate that the outer ring could be either due to an episodic outflow or that it traces the dust condensation rim in the outflow.
In both cases, we think that the outer ring is linked with the outflow and probably with the hourglass structure that we see at very large scales with ALMA.
To distinguish between these two interpretation we need to re-image this target with the VLTI to look for changes from epoch to epoch.

\section{An imaging Large Programme: INSPIRING}
\label{sec:INSPIRING}

These first two images, of IRAS08544-4431 and HD101584, show very different and complex morphologies that can be seen in the infrared at milliarcsecond scales as it was also indicated by the VLTI/PIONIER snapshot survey\cite{Kluska2019}.
To build a solid understanding of the very close environment of these binaries, we need to know wether those images are representative of this class of objects.
An imaging survey of these systems is perfect to provide such an understanding.
We have therefore initiated a Large Programme (PI: Kluska) that we named INterferometric Survey of Post-agb bInaries with their RING (INSPIRING) on the Very Large Telescope Interferometer (VLTI) using the PIONIER and GRAVITY instruments.
The main goal of this Large Programme is to image eleven targets to see if the picture we have from the images of IRAS08544-4431 and HD101584 is applicable to other such systems.
The scientific goals are the following:
\begin{itemize}
    \item reveal the 3-dimensional morphology of the disk inner rim and constrain disk/binary interactions
    \item obtain the absolute distances, luminosities, and masses of the systems by combining the radial velocity orbit with the spatially resolved angular position of the companion
    \item look for direct evidence for accretion from the circumbinary disk
    \item confirm the presence of a circum-secondary accretion disk in the sample
    \item study the origin of extended flux component 
    \item reveal a possible disk evolutionary sequence
\end{itemize}
Those goals will be achieved by performing image reconstruction of the continuum, locating the line absorption/emission from GRAVITY relative to the continuum derived from PIONIER and modelling the dataset together with the photometry with state-of-the-art 3-dimensional radiative transfer models\cite{Min2009}.

For the image reconstruction part, the procedure is the following: 
\begin{itemize}
    \item Perform an image reconstruction by subtracting a single star and fitting the chromatic parameters ($f^*_0$ and $d_\mathrm{env}$) in the process. In the case of a complete different morphology, perform classical image reconstruction.
    \item Perform the bootstrapping method to estimate the areas in the image where the emission is significant to be taken into account
    \item (optional) in case of detection of a second point source, perform another image reconstruction by subtracting the second point source as well. This step can be helped by model fitting to determine the flux and position of the secondary.
    \item Analyse the image geometry (e.g. is the ring centred on the primary? secondary? Can we deduce the centre of mass?)
    \item Use the obtained image to guide geometrical model fitting to the data and derive sizes, temperatures and flux ratios between the different components.
\end{itemize}
Such an approach should be sufficient to extract the most important features from the interferometric dataset and provide a direction for the interpretation and further modelling using more physical models.
A final consistency check should be done by generating synthetic dataset with noise from the final model and performed image reconstruction using the same hyper-parameters and compare to the original image.

Additionally, more advanced image reconstruction features can be tested on such a rich dataset such as data weighting (e.g., uniform or Briggs) that is used in radio interferometry.
Moreover, current image reconstruction algorithms use generic regularisation functions that are not representing our prior knowledge of the targets.
We, therefore, have developed a new image reconstruction algorithm based on deep learning techniques to learn a regularisation function from images generated by astrophysical models.
This is presented in this volume of conference proceeding by Claes et al.

\section{Conclusions}

Infrared interferometry brings important information to decipher binary evolution and the interaction of the binary with its close environment.
While first observations and snapshot surveys suggested a high complexity of the near-infrared emission geometry, only the first two images, on IRAS08544-4431 and HD101584, entirely revealed their morphology thanks to advances in image reconstruction techniques.
The SPARCO technique, originally developed for protoplanetary disks around young stars, showed to be very efficient to reveal the circumbinary environment of these targets.
Moreover, the bootstrapping technique to estimate the significant emission in the image was proven to be useful to guide the astrophysical interpretation and modelling of the images.
We also show the complementarity between ALMA and the VLTI.
The image of IRAS08544-4431 has particularly shaped our knowledge of these systems, by spatially resolving the different building blocks and allowing to analyse them one with respect to the other.
An imaging Large Programme is on-going to test this archetype and look for diversity in these intriguing and important to understand evolved binaries.
VLTI and CHARA are perfect infrared interferometric observatories to images these systems and progress on our knowledge.

\acknowledgments % equivalent to \section*{ACKNOWLEDGMENTS}       
J.K. acknowledges support from FWO under the senior postdoctoral fellowship (1281121N).
J.K. and H.V.W. acknowledge support from the research council of the KU Leuven under grant number C14/17/082.
J.A. and V.B. acknowledge support from the Spanish MICINN under project grants
AYA2016-789994-P (AxiN) and 
PID2019-105203GB-C21 (EVENTs / WEBULAE WEB).

% References
\bibliography{references} % bibliography data in report.bib
\bibliographystyle{spiebib} % makes bibtex use spiebib.bst

\end{document}